\documentclass[fleqn,twoside,10pt]{article}
\usepackage{amsmath}
\usepackage{espcrc2}
\usepackage{graphicx}
\usepackage{times}
\usepackage{psfrag}
\usepackage{mathbbol}

\newcommand{\fcaption}[1]{%
\vspace*{-6ex}%
\caption{#1}%
\vspace*{-3ex}}

\newcommand{\fig}[1]{fig.~\ref{#1}}

\newcommand{\hm}{{\hat{\mu}}}
\newcommand{\cp}{{\varphi}}
\newcommand{\contr}{{\mathcal{C}}}

\title{Vortices in the SU(2)--Higgs model\\
       Vortices and the covariant adjoint Laplacian}

\author{Roman Bertle\address[wien]{Atomic Institute of the Austrian Universities,
	Wiedner Hauptstra{\ss}e 8-10/142, A-1040 Vienna, Austria}%
	\thanks{Poster presented by R. Bertle.
		Supported in part by FWF 13997-TPH.},
	Manfried Faber\addressmark[wien],
	Albert Hirtl\addressmark[wien]}
       
\begin{document}

\begin{abstract}
\noindent\textbf{Vortices in the SU(2)--Higgs model}:
The presence of a fundamental Higgs in the SU(N)--Higgs model yields color screening at some finite distance.
Whereas the transition to the Higgs ``phase'' is accompanied by a suppression of projected center vortices, there is nearly no influence of color screening on the vortex properties in the confined ``phase''.
Hence the behavior of the Wilson loop can be described in both phases within the vortex picture of confinement. 

\noindent\textbf{Vortices and the covariant adjoint Laplacian}:
Laplacian center gauge is a method to localize center vortices in SU(N) gauge theory.
We show that the eigenvectors of the covariant adjoint Laplacian identify vortices for a special class of gauge field configurations.
However, for Monte Carlo generated configurations, modified approaches are required.
\vspace{1pc}
\end{abstract}

\maketitle

\section{VORTICES IN THE SU(2)--HIGGS MODEL}
Our action is
\begin{multline}
  S = S_{\text{W}} + \sum_x\bigl\{ \Phi^\dagger(x)\Phi(x) +
  \lambda\left[\Phi^\dagger(x)\Phi(x)-1\right]^2 \nonumber \\
  -\kappa\sum_{\mu}\left[
  \Phi^\dagger(x)U(x,\mu)\Phi(x+a\hat{\mu}) + \text{h.c.} \right] \bigr\},
\end{multline}
where $S_{\text{W}}$ is the Wilson action for SU(2) link variables $U$ and $\Phi$ is a scalar Higgs in the fundamental representation.
Due to color screening of fundamental charges, there is string breaking at some finite distance in the ``confined phase'', as in full QCD.
In the ``Higgs phase'' (at high values of $\kappa$), the string tension vanishes completely.
Note that these ``phases'' are analytically connected below some small inverse gauge coupling $\beta$.

The vortex picture explains the confinement properties of SU(N) gauge theory in terms of center vortices.
Thus it is interesting to study the influence of $\kappa$ on the properties of vortices.
We localize projected center vortices using the maximal center gauge.
All measurements have been done at $\lambda=0.5$.
\subsection{Confined ``phase''}
We compared $\kappa = 0$ with a $\kappa$ in the confined ``phase'' close to the ``phase transition''.
We find that the topological properties of vortices do not change with increasing $\kappa$, and the Higgs field does not lead to vortex depercolation, which would be one way to get screening in the vortex model.
This is consistent with the difficulties to detect string breaking with Wilson loops \cite{kn99b} -- in the vortex model, the infrared behavior of Wilson loops is determined by vortices.
Hence, whereas the picture is consistent, more subtle approaches are required to describe screening with vortices.
%
\subsection{``Phase transition''}
\begin{figure}
  \centering
  \psfrag{ti}[bc][bc]{$8^4$ lattice} \psfrag{KAPPA}{$\kappa$}
  \includegraphics[width=1.0\linewidth]{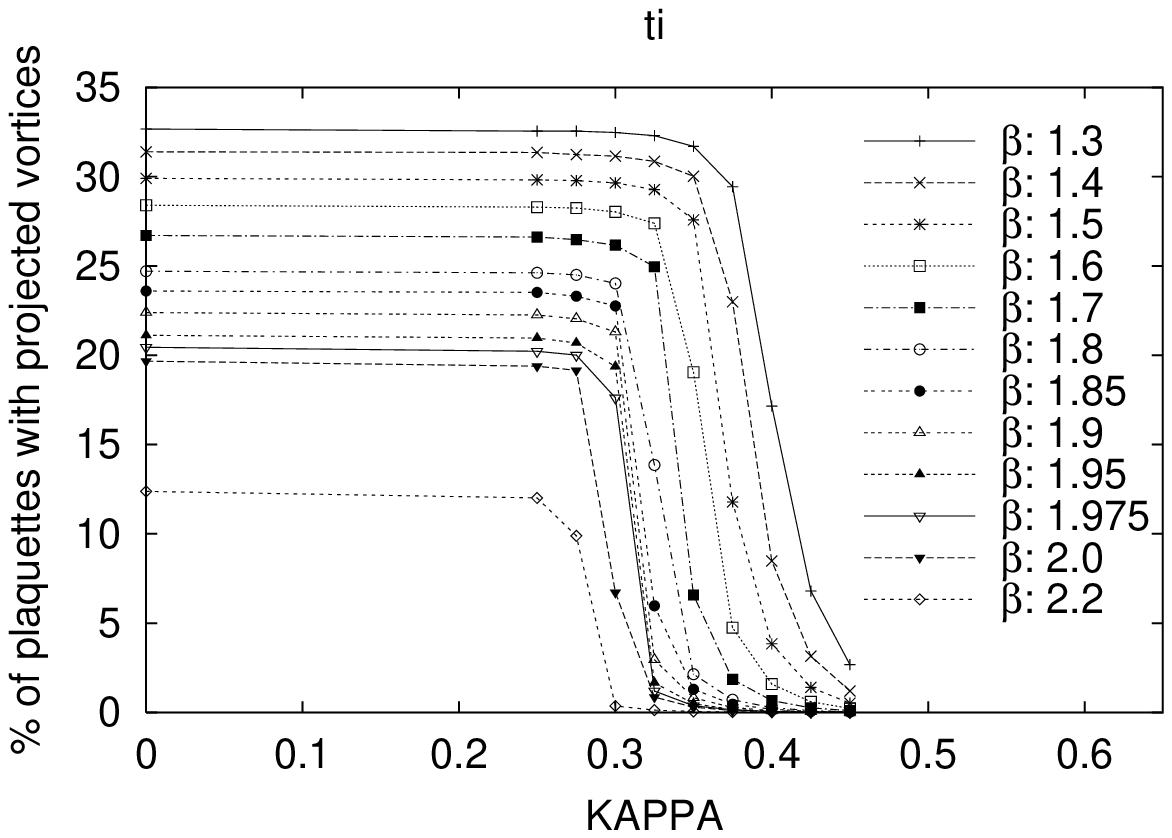}
  \fcaption{Density of vortex plaquettes.}
  \label{nplneg8}
\end{figure}

At the ``phase transition'' to the ``Higgs phase'', the string tension measured by Wilson loops disappears even for small distances.
This can be explained fully in terms of center vortices.
Vortices, which should be responsible for confinement, are strongly suppressed in the ``Higgs phase''.
The ``phase transition'' of the model is clearly reproduced even for a small lattice if we look at the vortex density plotted in \fig{nplneg8}.
\section{VORTICES AND THE COVARIANT ADJOINT LAPLACIAN}
 
\subsection{Laplacian center gauges}
\label{lcgtheo}
 
Laplacian center gauge (LCG) has been introduced \cite{fp00b} in order to overcome the Gribov problem of maximal center gauge (MCG).
In order to use a type of Laplacian center gauge, one introduces some scalar fields $\cp$ in the adjoint representation.
A choice of base vectors $e_i(x)$ in adjoint color space corresponds to a choice of some gauge for the gauge fields in the adjoint representation.
The covariant (adjoint) derivative is defined in such a base as
\begin{equation*}
  \nabla_\mu e_i(x) = U^j_{i\mu}(x) e_j(x+\hm) - e_i(x),
\end{equation*}
where $[U^j_i]_\mu(x)$ are the adjoint SU(2) link variables. 
For our scalar fields, we choose the eigenvectors of the adjoint covariant Laplacian: $ \Delta \cp^{(m)} = \lambda^{(m)} \cp^{(m)}$.
The eigenvectors of the lowest eigenvalues (called lowest eigenvectors henceforth) are the smoothest fields, i.e.\ their covariant derivatives are as small as possible.
Then orthonormalised base vectors $e_i$ are constructed which are close to these lowest eigenvectors.

For the Laplacian center gauge (LCG) \cite{fp00b}, one chooses at each lattices site $e_1(x)$ parallel to $\cp^{(1)}(x)$ and $e_2(x)$ parallel to the component $\cp^{\bot(2)}(x)$ of $\cp^{(2)}(x)$ orthogonal to $\cp^{(1)}(x)$.

In another recent proposal \cite{fgo01b}, a matrix from the first three eigenvectors $M_{ij}(x) = \cp^{(j)}_i(x)$ is built, from which the closest SO(3) gauge matrix is extracted using polar decomposition.

Because these base vectors are smooth, their covariant derivative is small.
Hence this is a gauge shifting the (adjoint) link variables $U^j_{i\mu}(x)$ close to unity, just as maximal center gauge does.
After performing such a gauge, it is possible to perform center projection in order to identify center vortices.

In contrast to MCG, the Laplacian gauges are not smooth in the very inside of a thick vortex.
This overcomes the continuum limit problem of MCG, and is related to gauge ambiguities.
%
\subsection{Gauge ambiguities}
\label{ambiguities}
For a trivial gauge field, i.e.\ a pure gauge, the lowest eigenvalue of the adjoint covariant Laplacian is three-fold degenerate.
The corresponding eigenvectors $\cp^{(i)}(x)$ are orthogonal at each lattice site and covariantly constant.
In this section, we assume that our gauge field deviates only by small fluctuations from a pure gauge.
In addition we insert some center vortices with finite, but small (compared to the outside) diameter -- they are well separated and do not overlap.
Outside of their core they are invisible for the adjoint Laplacian.
Thus the lowest eigenvectors are only slightly changed by them.

\begin{figure}
  \centering
  \psfrag{P}{$P$}\psfrag{Q}{$Q$}
  \psfrag{U}{$O(\phi)$}
  \psfrag{p}{$\cp$}
  \psfrag{Vortex}{Vortex}\psfrag{C}{$\contr$}
  \psfrag{ph}{$\phi$}\psfrag{x}{$x$}\psfrag{y}{$y$}
  \includegraphics[width=0.89\linewidth]{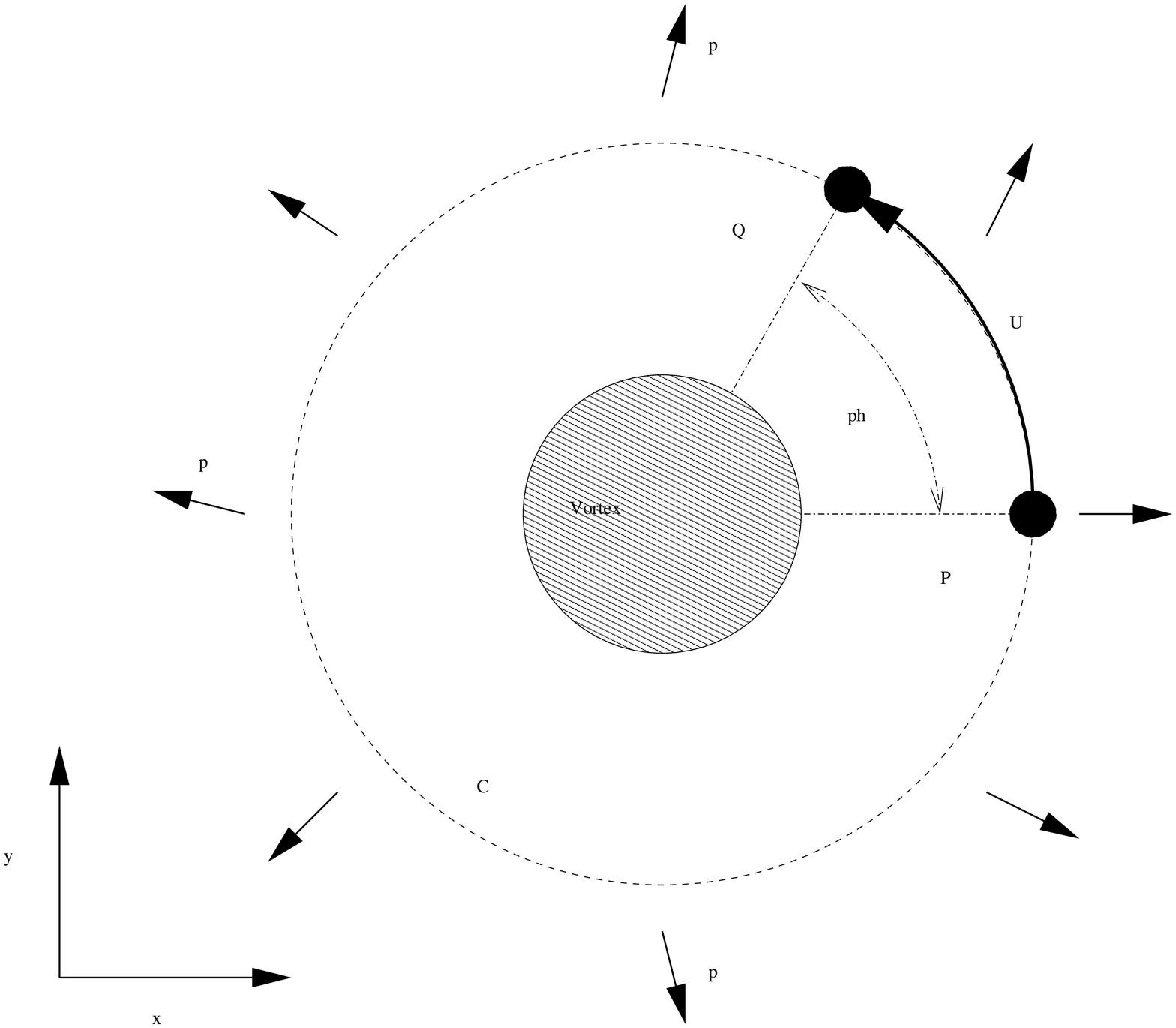}
  \vspace*{2ex}
  \fcaption{Simple vortex configuration.}
  \label{vort2d}
\end{figure}

Due to $\nabla_\mu \cp^{(i)}(x) \simeq 0$, in the configuration of \fig{vort2d}
the vectors on the circle $\contr$ are generated by $\cp^{(i)}(\phi)=O(\phi)\cp^{(i)}(P),\; O \in SO(3),\; \phi \in [0,2\pi)$.
This gives a map $\phi \rightarrow O(\phi)$ which is the nontrivial, noncontractible element of $\pi_1 (SO(3)) \hat{=} \mathbb{Z}_2$ because of the encircled center vortex.
Thus, in any smooth gauge one can see a rotation of $2\pi$ of the $\cp^{(i)}(x)$ around $\contr$.

If the vectors $\cp^{(i)}(x)$ would be smooth and linearly independent on the disk $D_2$ bordered by $\contr$, one could construct a smooth base $e_i(x)$ using orthogonalization.
This gives a smooth matrix field $O_{ij}(x)=e^{(k)}_i(x)\,{e^T}^{(k)}_j(P)$, $x \in D_2$.
But the map $D_2 \rightarrow SO(3)$ is contractible, in contradiction to the noncontractibility of $\contr \rightarrow SO(3)$ given above.
Thus the $\cp^{(i)}$ are linearly dependent somewhere inside $D_2$, which gives a gauge independent vortex localization.

We have looked at some configurations containing thick vortices inserted by hand.
In order to depict the rotation, we have selected a gauge smoothing the links in the plotted plane in \fig{eigenvec}.
The results agree with related investigations of Montero \cite{mo01a}.
\subsection{Spherical vortex}
%
%
We studied a spherical vortex that has the shape of the surface of a sphere of radius $R=6$.
It is contained in a 3d time-slice, the center of the sphere is at $x=y=z=t=10$.
The links in $t$- direction $U_t(r)$ vary from $-\mathbb{1}$ to $+\mathbb{1}$ between $r=R-d/2$ and $r=R+d/2$, with $d=1.9$.
The direction in color space $\vec{n}$ in which the links change is given by the radius vector $\vec{x}$ on the Euclidean lattice: we set $\vec{n}=\vec{x}/||\vec{x}||$.
Thus the link variables cover SU(2) completely.
 
\begin{figure}
  \centering
  \includegraphics[width=0.98\linewidth]{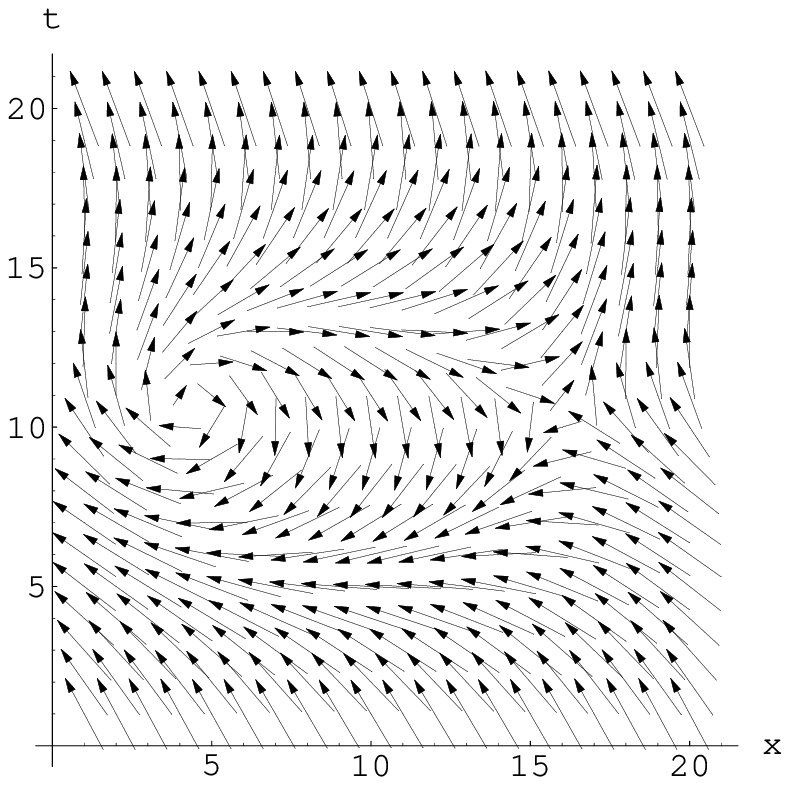}
  \fcaption{$\cp^{(1)}(x)$}
  \label{eigenvec}
\end{figure}

The expected rotation of the eigenvectors can be seen in \fig{eigenvec}.
The vector field $\cp^{(1)}(x)$ rotates around the two positions $t=10$, $x=4$ and $x=16$ where the vortex pierces the plotted plane (only two components of the vector field are drawn).

\begin{figure}
  \centering
  \includegraphics[width=0.98\linewidth]{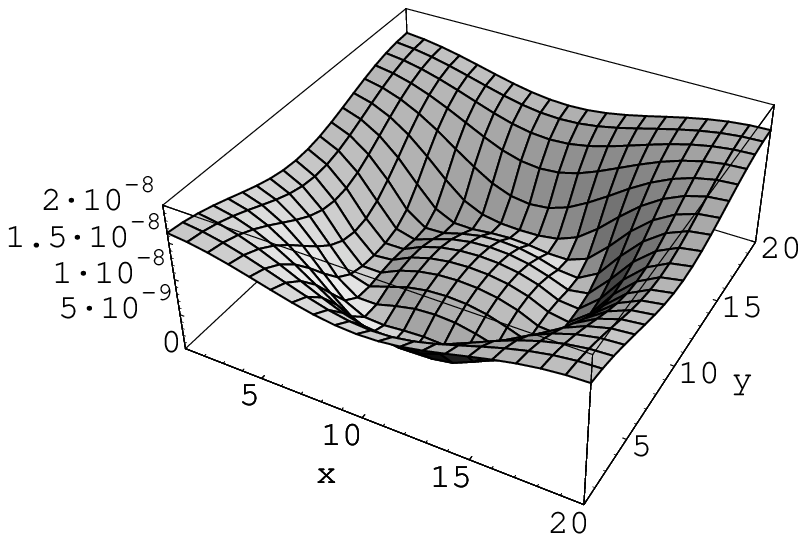}
  \fcaption{$\cp^{(1)}(x)\cdot(\cp^{(2)}(x)\times \cp^{(3)}(x))$}
  \label{spat}
\end{figure}

In order to detect the vortex position using the gauge ambiguity, we looked at the linear dependence of the vectors using the triple scalar product $\cp^{(1)}\cdot(\cp^{(2)} \times \cp^{(3)})$, which is the determinant of $M$.
In \fig{spat}, $\text{det}(M)$ approaches $0$ exactly at the position of the vortex.
The depicted plane cuts the vortex through a great circle which is clearly visible.
%
%
\subsection{Monte Carlo generated configurations}
For realistic SU(2) configurations, we failed to localize vortex positions from the rotation and the singularities of the lowest eigenvectors.
First, this is due to ultraviolet fluctuations which overshade the vortex structure.
Second, vortices are not isolated, Monte Carlo generated vortices even overlap.
As a result the components of the eigenvectors are suppressed in large regions.

This may also explain the P-vortex density in configurations after LCG and center projected configurations.
The vortex density does not scale, too many vortices are found at large $\beta$.

To conclude, whereas the method works fine for isolated vortices, for dynamical configurations modified approaches are required \cite{fgo01b}.
\subsection*{Acknowledgements}
We would like to thank Jeff Greensite and {\v S}tefan Olejn{\'{\i}}k for cooperation and interesting discussions.
\bibliographystyle{kphunsrteprint}
\bibliography{bertle}
\end{document}